\documentstyle[aps]{revtex}
\newif\iffinalplot

\finalplottrue
\begin{document}
\preprint{\small $ \begin{array}{r}
                   \rm RUB-TPII- 10/96\\
                   \rm hep-ph/9607219 \\
                   \rm June, 1996
                   \end{array} $}
\input prepictex
\input pictex
\input postpictex
\vfill\eject
\title{Phase transition and nucleon as soliton in the Nambu -- Jona-Lasinio 
model at finite temperature and density}
\author{Jochen Berger
\footnote{email:jochenb@hadron.tp2.ruhr-uni-bochum.de}, Christo V.~Christov
\footnote{email:christoc@neutron.tp2.ruhr-uni-bochum.de\\
Permanent address: Institute for nuclear research
and nuclear energy, Sofia 1784, Bulgaria}}
\bigskip
\bigskip
\address{
Institute for  Theoretical  Physics  II, Ruhr-University Bochum, \\
 D-44780 Bochum, Germany}
\bigskip
\bigskip
\bigskip
\maketitle
\tighten
\vskip0.5cm
\begin{abstract}
We study some bulk thermodynamical characteristics,  meson properties and
the nucleon as a baryon number one soliton in a hot quark matter in 
the NJL model as well as in a hot nucleon matter in a hybrid NJL model in  
which the Dirac sea of quarks is combined with a Fermi sea of nucleons. In 
both cases, working in mean-field approximation, we find a chiral 
phase transition from Goldstone to Wigner phase. At finite density the chiral 
order parameter and the constituent quark mass have a non-monotonic temperature
dependence - at finite temperatures not close to the critical one they are 
less affected than in the cold matter. Whereas the quark matter is rather 
soft against thermal fluctuations and the corresponding chiral phase 
transition is smooth, the nucleon matter is much stiffer and the chiral 
phase transition is very sharp. The thermodynamical variables show large 
discontinuities which is an indication for a first order phase transition.
We solve the $B=1$ solitonic sector of the NJL model in the presence of an 
external hot quark and nucleon medium. In the hot medium at intermediate 
temperature the soliton is more bound and less swelled 
than in the case of a cold matter. At some critical temperature, which for the 
nucleon matter coincides with the critical temperature for the chiral phase 
transition, we find no more a localized solution. According to this model 
scenario one should expect a sharp phase transition from the nucleon to 
the quark matter.  
\end{abstract}
{\sc Pacs} numbers: 12.39.Fe, 14.20.Dh, 14.40.Aq, 11.10.Wx
%
\section{Introduction}
%
At some finite density and/or temperature it is generally expected a
restoration of the chiral symmetry and a deconfinement, and hence a
change of the structure of the hadrons immersed in a hot and dense medium.
It is a topic of an increasing interest related to
the evolution of the early Universe as well as to the processes in the
interior of the neutron stars. Rather encouragingly,  
direct experimental studies of such phenomena in the relativistic heavy-ion 
reactions are also possible. Indeed, the lattice QCD calculations  suggest 
phase transitions concerning both chiral symmetry restoration and color 
deconfinement at relatively low temperatures 
(of about 150 - 160 MeV)~\cite{Fukugita88,Karsch95}. These
low temperatures are experimentally accessible even in the existing
accelerators and the corresponding heavy-ion experiments are already planned
at BNL and CERN~\cite{QMatter93}. From the side of theory the low critical 
values mean that we have to deal with non-perturbative phenomena. Since the 
analytical as well as the numerical (lattice QCD) methods are not developed 
enough to allow fully the solution of low-energy non-perturbative, especially 
if the baryons are involved, it is a motivation to apply
effective quark models. The Nambu - Jona-Lasinio (NJL) 
model~\cite{Nambu61} is the simplest purely quark model which incorporates 
the chiral symmetry and allows for its spontaneous 
breaking but it lacks confinement. It shows, however, a significant success in
description of meson structure as well as of the properties of the nucleon
as a non-topological soliton. Quite encouragingly,
assuming that the nuclear medium can approximately be replaced by a uniform
quark medium, the model describes (for review see~\cite{Hatsuda94}) a 
restoration of the chiral symmetry at finite temperature
and/or density in a quantitative agreement with the lattice QCD
calculations~\cite{Fukugita88,Karsch95} as well as with the chiral 
perturbation theory~\cite{Gasser87}. Following a scenario in which the 
restoration of the chiral symmetry in medium is considered as a relevant 
mechanism for a 
scale change which modify the hadron structure, the NJL model seems to 
provide an appropriate working scheme to study the hadron  
properties in hot medium. Assuming that the baryon medium can 
be approximated by a quark medium the bulk thermodynamical variables and 
medium modified meson properties has been repeatedly studied
\cite{Hatsuda94,Bernard87,Asakawa89,Christov91,Lutz92,Hufner94,Zhuang94,Rehberg95}.
The nucleon as a $B=1$ soliton in a cold quark medium has been 
already investigated~\cite{Christov93}. The results show that the nucleon 
as a soliton at finite density is less bound and increases in size, and at 
some critical value of about two times nuclear matter density  
the nucleon as a soliton does not exist. 

In the present paper our task is to study some bulk thermodynamical 
characteristics and meson properties  at finite temperature and density as 
well as the medium modification of the 
nucleon as a bound state of $N_c$ quarks coupled to the polarized medium 
(Fermi sea) and Dirac sea. 
In order to respect to some 
extent confinement we consider two different cases, namely a medium of 
nucleons, which is relevant before the phase transition, as well as a 
quark medium. In the former case, similar to the approach of 
Walecka~\cite{Serot86}, the nucleons are coupled directly to the meson fields. 
However, the NJL model differs from the Walecka approach in the 
meson fields, which appear basically as quark-antiquarks excitations, and 
also in the way
in which the chiral symmetry is dynamically broken. Such a model picture
has been considered by Jaminon et al.~\cite{Jaminon89} in the case of cold 
matter. 
Solving the $B=1$ solitonic sector of the NJL model in the 
presence of an external hot medium in a self-consistent 
way with the polarization of both Fermi and Dirac sea taken 
into account, we study the modification of the nucleon structure due to the 
medium. We work in mean-field approximation, 
which means that the meson quantum (loop) effects are not included in our 
considerations. Since the meson loop effects are 
dominant at low temperatures and vanishing density (pions are the lightest 
mode)~\cite{Hufner94,Zhuang94}, we restrict ourselves to consider mainly the 
case of finite density and large enough temperatures where the nucleon
and quark degrees of freedom are most relevant. This case is 
also of interest for the relativistic heavy-ion reactions.
\section{NJL model at finite temperature and density}

We use the Nambu--Jona-Lasinio (NJL) model~\cite{Nambu61} which is the
simplest purely quark model
describing spontaneous chiral symmetry breakdown. The SU(2)-version of the
NJL lagrangian contains chirally invariant local scalar and pseudoscalar
four-quark interaction:
\begin{equation}
{\cal L} = \bar \Psi \, (i  \partial\hskip-6pt/ - m_0)\, \Psi +
{G\over 2} \, [(\bar \Psi \Psi )^2 + (\bar \Psi i \vec\tau
\gamma_{5} \Psi )^2\,] \ ,
\label{NJLLAG} \end{equation}
where $\Psi$ is the quark field, $G$ is the coupling constant,
$\vec\tau$ are the Pauli matrices in the isospin space and
$m_0$ is the current mass taken equal for both $up$ and $down$
quarks. Applying the well known bosonization procedure~\cite{Eguchi74}
the NJL model is expressed in terms of auxiliary meson fields
$\sigma, {\vec\pi}$:
\begin{equation}
{\cal L} = \bar \Psi \  ( \ i \partial\hskip-6pt/ -
\sigma - i\, {\vec\pi\cdot\vec\tau}\, \gamma_5 \ ) \, \Psi -
{1\over {2G}} (\sigma^2 + {\vec\pi}^2)+{m_0\over G}\sigma\,.
\label{NEWLAG} \end{equation}

Using functional integral technique the quantized theory at finite temperature
and density can be written in terms of the corresponding euclidean grand
canonical partition function~\cite{Bernard74}:
\begin{equation}
Z=\mbox{Tr}\, \exp\Bigl\{-\beta(H-\mu N)\Bigr\}={1\over Z_0}\int {\cal D}
\Psi{\cal D}
\Psi^\dagger\, \exp\Bigl\{\int_0^\beta\mbox{d}\tau\int_V 
\mbox{d}^3x ({\cal L}-\Psi^\dagger\mu\Psi)\Bigr\}\,,
\label{partfuncion}
\end{equation}
where $V$ is the volume of the system, $\beta$ is the inverse temperature and
$\mu$ is the chemical potential.
The integration over the quarks can be done exactly, whereas for the
integration over the mesons we use a large $N_c$ saddle-point (mean-field) 
approximation. It means that the meson fields are treated classically -- no
meson loops are taken into account. Following refs.~\cite{Bernard74}
we replace the integration over the imaginary time by a sum over fermionic
Matsubara frequencies $k_0\rightarrow (2n+1)\pi/\beta$. Finally we get for the
effective action
\begin{eqnarray}
S_{eff}(\mu,\beta)=-\ln Z&=&-\beta V N_c\sum_\alpha
\Biggl\{\frac 12(\epsilon_\alpha-\mu)+\frac 1\beta
\ln\Bigl[1+\mbox{e}^{-(\epsilon_\alpha-\mu)\beta}\Bigr]\Biggr\}\nonumber\\
&+& \beta\int_V \mbox{d}^3x\Bigl[{1\over 2G}(\sigma^2+{\vec \pi}^2)-
{m_0\over G}\sigma\Bigr]\,.
\label{Seff}
\end{eqnarray}
The energies $\epsilon_\alpha$ are eigenvalues of the one-particle 
hamiltonian $h$:
\begin{equation}
h\Phi_n\equiv\Bigl[\frac {\vec\alpha\cdot\vec\nabla}i+\gamma_0 (\sigma+
i\gamma_5\vec\pi\cdot\vec\tau)\Bigr]
\Phi_n = \epsilon_n \Phi_n\,
\label{dirac}
\end{equation}
and $\Phi_n$ are eigenfunctions.
The saddle-point solution makes the
effective action stationary
\begin{equation}
{\partial S_{eff}\over\partial\sigma}\Biggr\vert_{\sigma_c}={\partial
S_{eff}\over\partial\vec\pi}\Biggr\vert_{\vec\pi_c}=0\,,
\label{statcond}
\end{equation}
with the number of particles $N$ in the volume $V$ 
\begin{equation}
N=-{1\over \beta}
{\partial S_{eff}\over\partial\mu}\Biggr\vert_{\sigma_c,\vec\pi_c}\,.
\label{mu}
\end{equation}
kept fixed.
Here $\sigma_c$ and $\vec\pi_c$ are the ``classical'' values
of the meson fields and $\mu$ is the chemical potential 
related to the number of particles $N$.

The thermodynamical characteristics of a many-body
system are specified by the thermodynamical potential
\begin{equation}
\Omega(\mu,\beta)\equiv \frac {S_{eff}(\mu,\beta)}{\beta V}
\label{omega}
\end{equation}
It should be noticed that the saddle-point solution ($\sigma_c,\vec\pi_c$) 
minimizes not $\Omega$ but the Helmholtz free energy
\begin{equation}
F=\Omega-\mu{\partial\Omega\over\partial\mu}\,,
\label{freeenergy}
\end{equation}
with a constraint (\ref{mu}) and $\mu$ playing a role
of a Lagrange multiplier.

In the mean-field approximation (leading order in $1/N_c$) the inverse 
meson propagator is given by the second variation of the 
effective action at the stationary point $\sigma_c,\vec\pi_c$:
\begin{equation}
K_{\phi}^{-1}(x-y)={\partial^2 S_{eff}\over \partial 
\phi(x)\partial\phi(y)}\Biggr\vert_{\phi_c}\,,
\label{mesonprop}
\end{equation}
where $\phi$ stands for both meson fields.
The on-shell meson masses correspond to the poles of the meson propagator
\begin{equation}
K_{\phi}^{-1}\biggl(q_0^2=-m^2_\phi\biggr)=0\,
\label{mesonmass}
\end{equation}
at $\vec q=0$ 
and the physical quark meson coupling constants are given by the residue of
the propagator at the pole
\begin{equation}
g_{\phi}^2=\lim\limits_{q^2\rightarrow -m^2_\phi}(q^2+m^2_\phi)K_{\phi}
(q^2)\,.
\label{couplconst}
\end{equation}

Due to the local four-fermion interaction the
lagrangian (\ref{NJLLAG}) is
not renormalizable and a regularization procedure
with an appropriate
cut-off $\Lambda$ is needed to make the effective
action finite.
Actually only the part of the effective action $S_{eff}(\mu=0,\beta=\infty)$,
coming from the Dirac sea (negative-energy part of the spectrum),
is divergent and we regularize it using the proper-time regularization
\begin{equation}
   \mbox{Tr} \ln \hat A \rightarrow -\mbox{Tr}\int\limits_{\Lambda^{-2}}^\infty
{\mbox{d}s\over s}\mbox{e}^{-s \hat A}\,.
\label{proptime}
\end{equation}
It is easy to check that the difference
$S_{eff}(\mu,\beta)-S_{eff}(\mu=0,\beta=\infty)$
is finite and does not need any regularization. Such a scheme is consistent 
with the usual regularization used in the solitonic calculations in the
NJL model (see \cite{Christov96} and the references therein) in 
which only the divergent part of the effective action is regularized, 
whereas the finite valence part is not affected by the 
regularization. Moreover, any regularization
of the medium part would suppress the temperature 
effects~\cite{Christov91}, since
the positive part of the spectrum would be affected by the cutoff as well.
Thus regularizing only the divergent Dirac sea part, we obtain the effective 
action in the form
\begin{eqnarray}
S_{eff}(\mu,\beta) &=& \beta V N_c \Bigl\{\sum_{\epsilon_\alpha<0}\bigl[
R^\Lambda_{3/2}(\epsilon_\alpha)+(\mu-\epsilon_\alpha)\bigr]
+ {1\over \beta} \sum_\lambda \ln\Bigl[1+\mbox{e}^{-\beta(\epsilon_\alpha-\mu)}
\Bigr]\Bigr\}\nonumber\\ 
&+& \beta \int_V \mbox{d}^3x 
\Biggl[\frac{1}{2G}\Bigl(\sigma^2
+\vec{\pi}^2\Bigr)-\frac{m_0}{G} \sigma \Biggr]
\label{regseff}
\end{eqnarray}
where the proper-time regularization function is given by
\begin{equation}
R^\Lambda_{\alpha}(\epsilon) = {1\over\sqrt{4\pi}}\int_{\Lambda^{-2}}^\infty
\frac{d\tau}{\tau^\alpha} \mbox{e}^{-\tau\epsilon^2}\,.
\label{regener}
\end{equation}
Since the Dirac sea does not contribute to the baryon number, 
expression (\ref{mu}) is finite and does not include any
regularization.

\section{Fixing the model parameters}

In the vacuum ($\mu=0,T=0$) the stationary conditions (\ref{statcond}) lead to
a translationary invariant solution
\begin{equation}
\sigma_c=M_0 \qquad \mbox{and}\qquad  \vec\pi_c=0\,,
\label{mesconfv}
\end{equation}
which breaks the chiral symmetry. Due to the non-zero vacuum expectation value
$\sigma_c$ the quarks acquire a constituent mass $M_0$. The latter is a
solution
of the gap equation (stationary condition for the sigma field) which in the
case of proper-time regularization has the form
\begin{equation}
M_0=\,G\,M_0{N_c\over2 \pi^2}\int_{\Lambda^{-2}}^\infty{\mbox{d}s\over s^2}
\mbox{e}^{-sM^2}+m_0\equiv m_0-G< \bar \Psi \Psi >\,
\label{gap}
\end{equation}
with $< \bar \Psi \Psi >$ being the quark condensate.
Using (\ref{mesonprop}) it is straightforward (see
ref.~\cite{Jaminon89}) to evaluate the meson propagator:
\begin{equation}
K_\phi(q^2)={1\over Z_p(q^2)}{1\over q^2+\delta_{\phi\sigma}4M_0^2
+{m_0\over GM_0  Z_p(q^2)}}\,.
\end{equation}
The function $Z_p(Q^2)$ corresponds to a quark loop with two 
pseudoscalar-isovector insertions ($i\gamma_5\tau_a$). Here we present
only the proper-time regularized expression for it:
\begin{equation}
Z_p(Q^2)={N_C \over 4\pi^2}\int\limits_0^1\mbox{d} u
\int\limits_{\Lambda^{-2}}^\infty {\mbox{d} s\over s}
\mbox{e}^{-s\biggl[M_0^2+{q^2\over 4}(1-u^2)\biggr]}=
{N_C \over 4\pi^2}\int\limits_0^1\mbox{d} u 
\Gamma\Bigl(0,{M_0^2+{q^2\over 4}(1-u^2)\over \Lambda^{-2}}\Bigr) \,,
\label{Zpq2}
\end{equation}
with $\Gamma(0,x)$ being the incomplete gamma function.

We fix the parameters of the model, namely the current mass $m_0$, the
cutoff $\Lambda$ and the coupling constant $G$ in the vacuum sector
reproducing the physical pion mass $m_\pi=140$ MeV and the pion decay
constant $f_\pi=93$ MeV. In fact, it leads to the Goldberger-Treiman (GT) 
relation on the quark level
\begin{equation}
M_0=g_\pi f_\pi\,,
\label{GT}
\end{equation}
and one also recovers the Gell-Mann - Oakes - Renner (GMOR) relation:
\begin{equation}
m_\pi^2 f_\pi^2=-m_0< \bar \Psi \Psi >+O(m_0^2)\,.
\label{GMOR}
\end{equation}
As usual, the last model parameter, the coupling constant $G$, is
eliminated in favor of the constituent mass $M_0$ using the gap equation 
(\ref{gap}). In fact, a
value around $M_0=420$ MeV is needed to describe properly the nucleon 
properties~\cite{Christov96} and we will use this value in our computations. 
The corresponding value of the proper-time cutoff is $\Lambda=640$ MeV.

\section{Phase transition and meson properties at finite temperature and 
density}

In this section we investigate the translationary invariant medium solution 
at finite temperature $T$ and density $\rho$. For the model parameters 
$m_0$, $\Lambda$ and $M_0$ we use the values fixed in the vacuum.

\subsection{Quark medium}

In this subsection we consider the medium as a Fermi sea of quarks.
Solving (\ref{statcond}) together with the number of particles $N$ in the 
volume $V$ (\ref{mu}) kept fixed, we get the the medium solution in the form
\begin{equation}
\sigma_c=M \qquad \mbox{and}\qquad  \vec\pi_c=0\,.
\label{medmesconf}
\end{equation}
As in the vacuum case, only the ``classical'' value of the scalar
field can be non-zero and we call it the constituent quark mass $M$ in
medium. This mass $M$ is a solution of the equation of motion  the scalar field
\begin{figure}[th]
\input{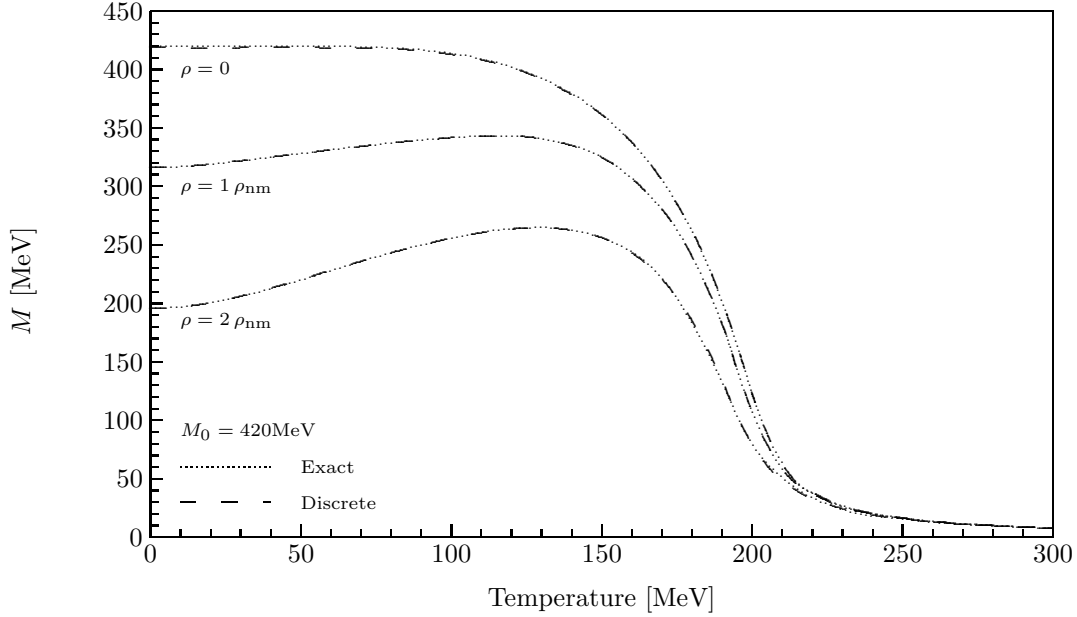}
\caption{Constituent quark mass $M$ as a function of temperature for
different densities. At vanishing both temperature and density it is
fixed to $M_0 = 420$ MeV. The results with plane-wave basis (dotted
lines) are compared to those of the finite-discrete basis (dashed lines).}
\label{Figr1}
\end{figure}
\begin{equation}
M=-\,G\,N_c\Biggl[\sum_{\epsilon_n<0}\bar\Phi_n\Phi_n 
R_{1/2}^\Lambda(\epsilon_n)\epsilon_n
+\sum_{\epsilon_n>0}{\bar\Phi_n\Phi_n
\over 1+\mbox{e}^{(\epsilon_n-\mu)\beta}}
-\sum_{\epsilon_n<0}{\bar\Phi_n\Phi_n
\over 1+\mbox{e}^{-(\epsilon_n-\mu)\beta}}\Biggr]+m_0\,,
\label{medgap}
\end{equation}
with the proper-time regulator $R_{1/2}^\Lambda(\epsilon_n)$ in the form
(\ref{regener}). The thermodynamical potential 
$\Omega$ is given by (\ref{omega}):
\begin{eqnarray}
\Omega(\mu,\beta) &=& N_c\sum_{\epsilon_n<0} 
\Bigl[R^\Lambda_{3/2}(\epsilon_n)-R^\Lambda_{3/2}
(\epsilon_n^0)+(\mu-\epsilon_n)\Bigr]
-{N_c\over \beta} \sum_n \ln\Bigl[1+\mbox{e}^{-\beta(\epsilon_n-\mu)}
\Bigr]\Biggr\} \nonumber\\
&+& {1\over V} \int_V \mbox{d}^3x 
\Biggl[\frac{1}{2G} (M^2-M_0^2)-\frac{m_0}{G} (M-M_0) \Biggr]\,.
\label{regomegam}
\end{eqnarray}
Here, we subtract the vacuum value of the thermodynamical potential 
$\Omega(0,0)$. The energies $\epsilon_n^0$ are the 
eigenvalues of the hamiltonian $h$ for the vacuum meson configuration 
(\ref{mesconfv}). Using (\ref{mu}) the baryon density $\rho_B$ can be written 
as 
\begin{equation}
\rho_B =-{1\over N_c}{\partial \Omega\over \partial \mu}
=\sum_{\epsilon_n>0}
{1\over 1+\mbox{e}^{(\epsilon_n-\mu)\beta}}
-\sum_{\epsilon_n<0}{1\over 1+\mbox{e}^{-(\epsilon_n-\mu)\beta}}\,,
\label{bardensity}
\end{equation}
and the free energy (\ref{freeenergy}) is given by
\begin{equation}
F(\mu,\beta)=\Omega(\mu,\beta)+N_c\mu\rho_B\,.
\label{freeenergym}
\end{equation}

In the calculations we use a plane wave basis as well as a
quasi-discrete basis and numerical method of Ripka and
Kahana~\cite{Kahana84} for solving the
eigenvalue problem (\ref{dirac}) by putting the system in a
spherical box  of a large radius $D$. The basis is 
made discrete by imposing a boundary condition at $D$. It is also made finite 
by introducing a numerical cut-off $k_{\rm max}$ for the
momenta of the basis states. Obviously both parameters are technical
and the results do not depend on their particular values. The
typical values which we use, are $D\approx 18/M_0$ and $k_{\rm max}
\approx 8 M_0$. The volume $V$ is taken to be a bit 
smaller than the box in order to avoid finite-size-box effects and 
the baryon number is given by 
\begin{equation}
N_B=\rho_B V\,.
\label{barnumber}
\end{equation}

\begin{figure}[thb]
\input{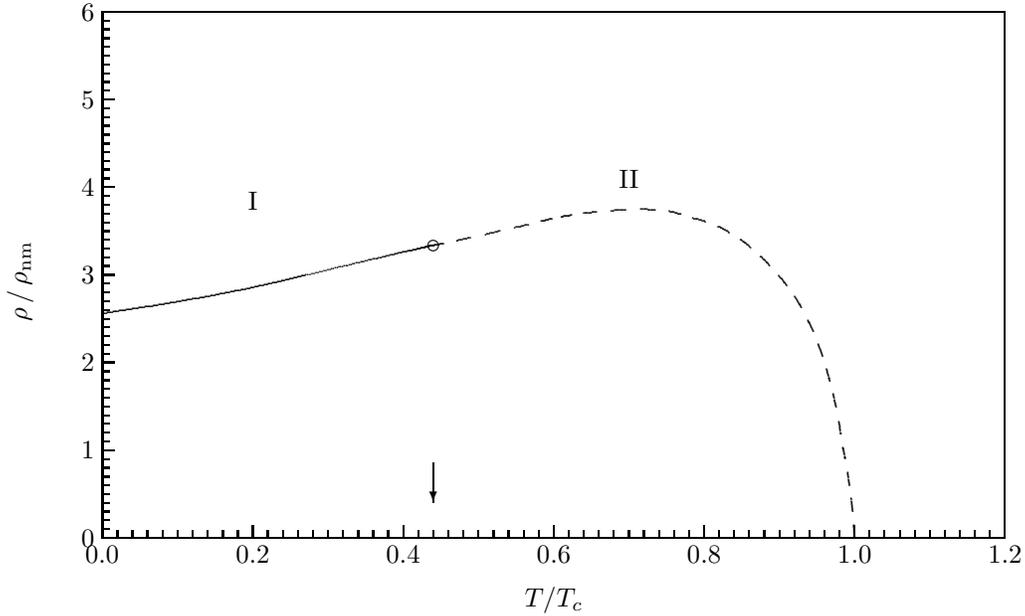}
\caption{$\rho - T$ critical phase diagram. The lines separates the
Goldstone phase with the chiral symmetry breaking from the Wigner
phase where the chiral symmetry is restored. The solid line shows the
critical values for which there is a first order transition, whereas
the dashed line corresponds to the second order one. The arrow shows
the temperature at which the order is changed.}
\label{Figr3}
\end{figure}

The results for the constituent mass $M$ as a function of temperature
for different densities are presented in
fig.~\ref{Figr1} for both the plane wave and the quasi-discrete basis.
As can be seen the quasi-discrete basis provides a quite good description of 
the continuum. 
\begin{figure}[th]
\input{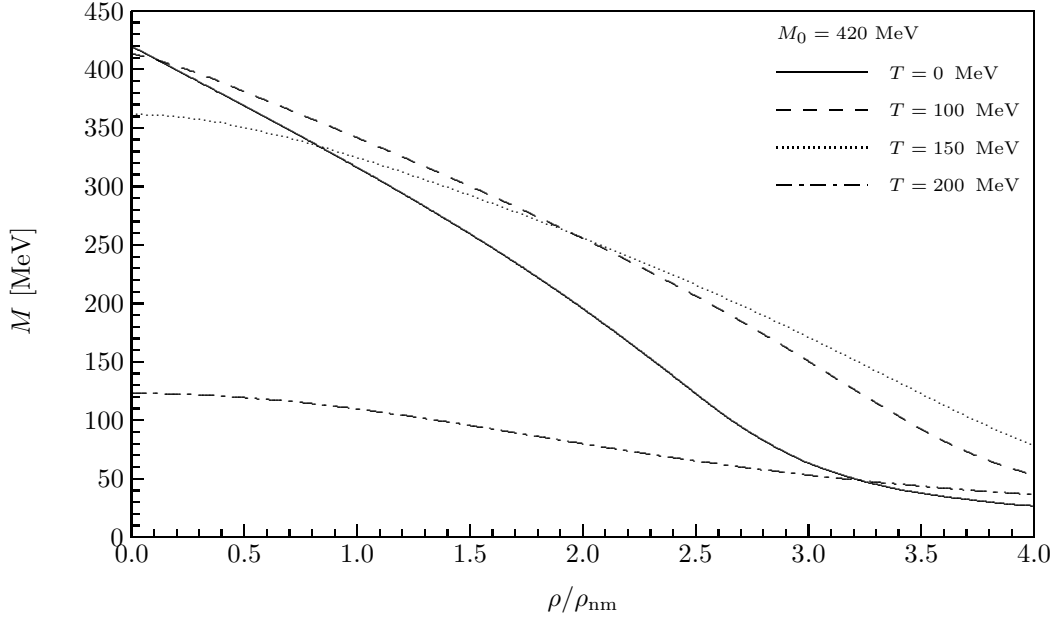}
\caption{Constituent quark mass $M$ as a function of density for
different temperatures.}
\label{Figr2}
\end{figure}
At finite density the constituent mass (or the chiral oder 
parameter $<\bar \Psi \Psi>$) is a non-monotonic 
function of temperature. It means that in 
a hot matter $M$ is less affected than in the case of cold matter. 
This also can be seen on fig.~\ref{Figr2} where $M$ is presented as function 
of density for different temperatures. At some critical values of temperature 
and/or density 
the constituent mass $M$ is reduced to the current mass $m_0$ (the quark 
condensate $<\bar \Psi \Psi>$ vanishes) which is an indication for a a 
transition to the Wigner phase where
the chiral symmetry is not spontaneously broken. The 
corresponding equation of state (pressure versus density) can be found in 
ref.~\cite{Christov91}. The critical $\rho - T$ chiral phase transition 
diagram is 
shown in fig.~\ref{Figr3}. At lower temperature values $T< 90$ MeV we have a 
clear first order transition which change to a second order one at higher 
temperatures. The critical value of the temperature is close to 200 MeV and 
for the density is about $2.5\ \rho_{nm}$ (nuclear matter density 
$\rho_{nm}=0.16\ \mbox{fm}^{-3}$). 

\begin{figure}[thb]
\input{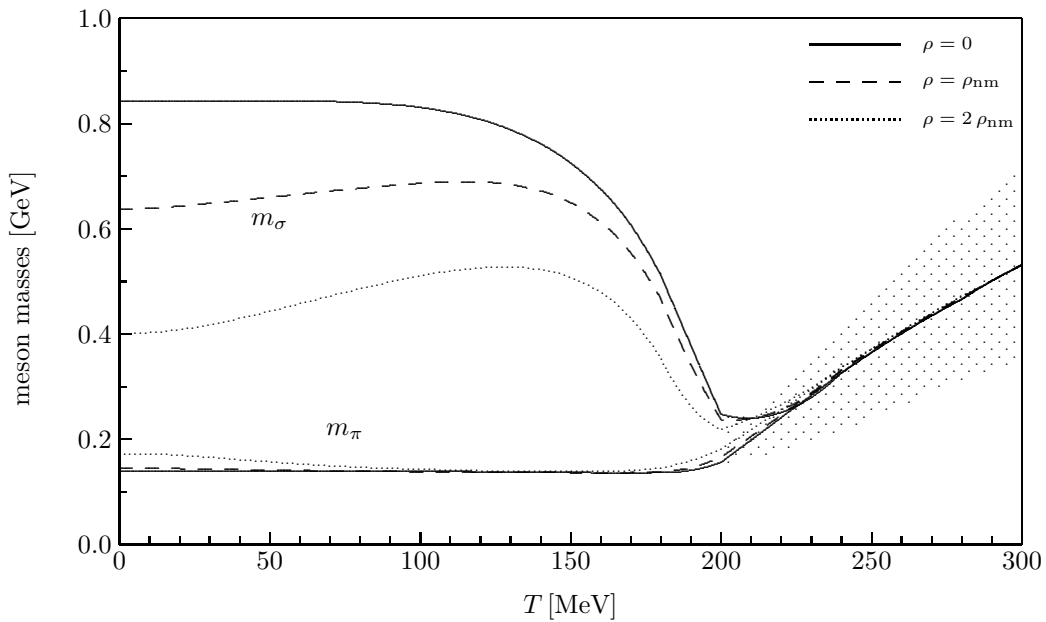}
\caption{Meson masses in the quark medium as a function of temperature for
different densities. The shadowed area shows the width of the meson
resonances.} 
\label{Figr4}
\end{figure}

\begin{figure}[thb]
\input{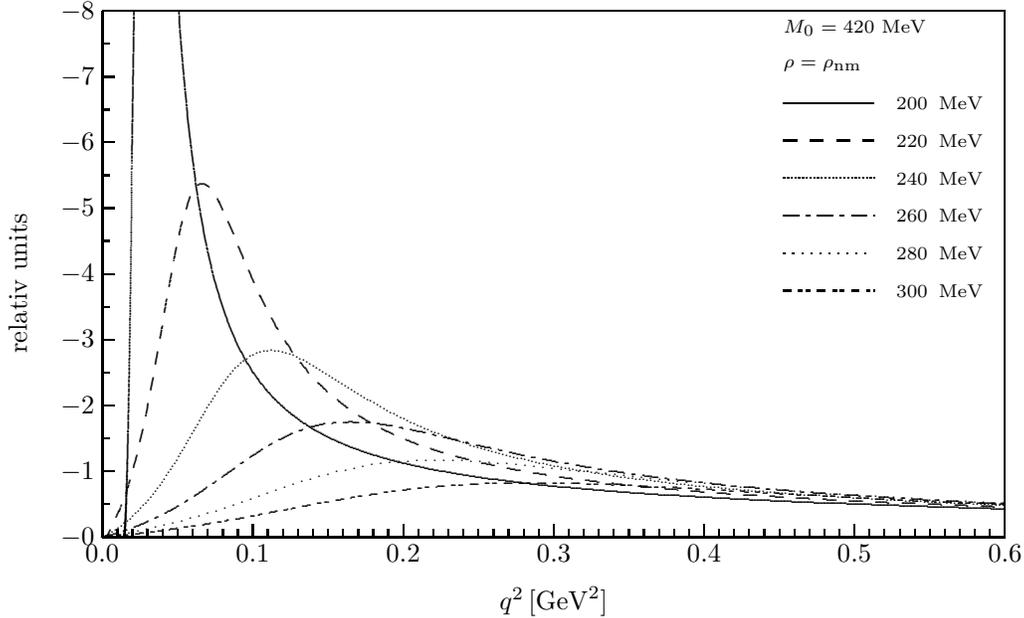}
\caption{Imaginary part of the pion propagator (spectral density) for
nuclear matter density and different temperatures.}
\label{Figr5}
\end{figure}

The meson masses in medium are defined as the lowest zero solution of the 
inverse meson propagators
\begin{equation}
K^{-1}_\phi(q_0^2=-m_\phi^2)=0\,,
\label{medmesmass}
\end{equation}
at $\vec q=0$. 
The latter is given by the second variation of the effective action 
with respect to the meson fields
\begin{eqnarray}
K^{-1}_\phi(q_0^2)&=&\Bigl({q_0^2\over 4}+\delta_{\phi\sigma}M^2\Bigr)
\Biggl\{ Z_p(q_0^2)-4N_c{\cal P}\int {\mbox{d}^3k\over 
(2\pi)^3} {1\over \epsilon_k^2+{q_0^2\over 4}}\Biggl[{1\over 
1+\mbox{e}^{\beta(\epsilon_k-\mu)}}
+{1\over 1+\mbox{e}^{-\beta(\epsilon_k+\mu)}}\Biggr]\Biggr\}\nonumber\\
&+&{m_0\over G M}\,.
\label{invprop}
\end{eqnarray}
In the derivation of (\ref{invprop}) we use the plane wave basis and 
eq.(\ref{medmesconf}). ${\cal P}$ means principle value. From 
(\ref{invprop}) it is clear 
that in the Goldstone phase the scalar and pseudoscalar mesons differ in 
mass. In the chiral limit the pions remain massless Goldstone bosons also in 
the medium, whereas the sigma
mass follows the constituent quark mass. Both GMOR and GT relations are 
also valid in medium. Since in the Wigner phase the constituent mass $M$ 
vanishes up to the current mass $m_0$, the mesons become degenerated in 
mass and appear as parity-doubled mesons. In fact, in the Wigner phase due 
to the Fermi sea contribution, even in the chiral limit $m_0=0$
the mesons have a non-zero mass, which increases with temperature (density). 
It is illustrated in fig.~\ref{Figr4} where our results for the meson masses 
in medium are shown  as a function of temperature at three different densities.
Similar results are obtained by Zhuang et al.~\cite{Zhuang94} in the case of 
finite temperature and vanishing chemical potential. 
Since the NJL model lacks confinement there is a 
coupling to the $\bar qq$ continuum. In the Goldstone phase this coupling is 
unphysical but it does not affect much the structure of the scalar and 
pseudoscalar mesons -- for the pion we have a $\delta$-peak far from a very 
weak $\bar qq$ continuum and for the sigma meson there is a
very sharp peak at the $\bar qq$ threshold ($4M^2$). The reason is that the 
pions play the role of Goldstone bosons and the sigma mesons are their chiral 
partner. In the Wigner phase, however, after the deconfinement 
transition (suggested by the lattice results) the $\bar qq$ channel is 
physically open. The meson states are not more stable. 
They appear as resonances in the $\bar qq$ continuum and
the poles of the meson propagators are complex. In this case one 
can identify the meson resonances with the peaks in the spectral density  
(imaginary part of the meson propagator).  The spectral density is shown on 
fig.~\ref{Figr5} for 
some different temperatures above the critical one. As can be seen,  
there is indeed a clear resonance structure. Close to the critical 
$T_{cr}\approx 200$ MeV the peak is rather narrow. With the increasing 
temperature the resonance structures become weaker
and at around 300 MeV is almost washed out. The position of the peak moves 
to higher energies and its width 
increases as well which implies that the mesons in the Wigner phase are rather 
soft modes than elementary excitations~\cite{Hatsuda94}. It is also 
interesting to notice that in this phase the meson masses show almost no 
density dependence. To conclude, according to the above model picture, 
at higher temperatures one might expect a smooth dehadronization transition.
In this range, not the hadron but the quark and gluon degrees 
of freedom are expected to be relevant. Apparently, 
the missing explicit gluonic degrees of freedom in the NJL model limits its 
applicability at high temperatures.      
\subsection{Medium of nucleons}

In the previous subsection  we have considered the quark medium. 
In this case at finite temperature the single 
quarks from the Dirac sea are allowed be excited and occupy the levels in 
positive part of the spectrum (Fermi sea) leaving holes (antiquarks) in the
Dirac sea. Apparently, because of confinement, at $T$ and $\rho$ below
the critical values for the chiral and deconfinement transitions it is  
forbidden. As an alternative to this 
picture in this subsection we use a hybrid model in which instead of quarks we 
consider a Fermi sea of nucleons. The Dirac sea consists of quarks and it 
determines the vacuum sector. In this model the mesons are still  
$\bar qq$ excitations but they are also directly coupled to the nucleons 
of the Fermi sea. 
Such a model have been proposed and used by Jaminon et al.~\cite{Jaminon89} to 
study the the chiral symmetry restoration in a cold nucleon medium. At finite 
temperature only nucleons (colorless $N_c$ quark clusters) can be removed from 
the Dirac sea and occupy the levels in the Fermi sea. In this model picture, 
removing nucleons from the Dirac to the Fermi sea,  
one creates antinucleons. Apparently, it is physically reasonable to apply
this hybrid model only for temperature and density below 
their critical values. As we see later, almost simultaneously with the chiral 
phase transition there is a delocalization of the soliton in the model, which 
means that after the phase transition it is consistent to 
consider a quark medium rather than the nucleon one.  

Since in the present scheme the contribution of the Dirac sea  is separated 
from the one of the Fermi sea, it is straightforward to write down the 
medium contribution to the thermodynamical potential (effective 
action) in terms of nucleons:
\begin{figure}[thb]
\input{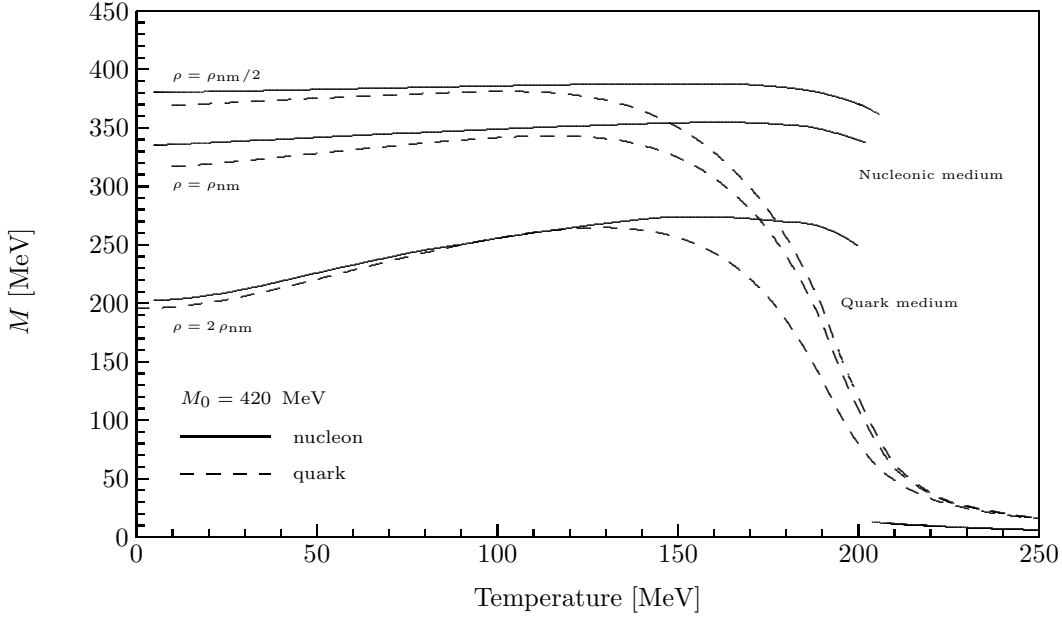}
\caption{Constituent quark mass $M$ as a function of temperature for
different densities in the quark (dashed lines) and nucleon medium
(solid lines).}
\label{Figr6}
\end{figure}
\begin{equation}
\Omega^N_{med}=\sum_{\epsilon_n^N<0} 
(\mu^N-\epsilon_n^N) - {1\over \beta} \sum_n 
\ln\Bigl[1+\mbox{e}^{-\beta(\epsilon_n^N-\mu^N)}
\Bigr] \,.
\label{nucmedium}
\end{equation}
The energies $\epsilon_n^N$ are the solutions of the Dirac equation
\begin{equation}
h_N\Phi_n^N\equiv\Bigl[\frac {\vec \alpha\cdot\vec\nabla}i+\beta g_N(\sigma+
i\gamma_5\vec\pi\cdot\vec\tau)\Bigr]
\Phi_n^N = \epsilon_n^N \Phi_n^N\,.
\label{ndirac}
\end{equation}
The meson fields are coupled to the nucleons with a 
coupling constant $g_N$ which relates the nucleon mass to the nonzero 
expectation value of the scalar meson field (constituent quarks mass $M_0$) in 
vacuum
\begin{equation}
M_N=g_N M_0\,.
\label{nucmassv}
\end{equation}
As in the quark spectrum, there is a gap of $2M_N$ in the nucleon spectrum 
which separates the negative part of the spectrum 
from the positive one. In the nucleon medium, we also assume that 
$g_N$ remains unchanged which means that a relation similar to 
(\ref{nucmassv}) is valid also in medium. Since we are able to 
calculate the mass of the nucleon as a soliton, later we will check this 
approximation. The chemical potential $\mu^N$ is fixed by the baryon density
\begin{equation}
\rho_B =\sum_{\epsilon_n^N>0}{1\over 1+\mbox{e}^{(\epsilon_n^N-\mu^N)
\beta}}-\sum_{\epsilon_n<0}{1\over 
1+\mbox{e}^{-(\epsilon_n^N-\mu^N)\beta}}\,,
\label{bardensnucl}
\end{equation}
(or equivalently by the baryon number (\ref{barnumber})).
It is interesting to make a parallel with the approach of Walecka et 
al.~\cite{Serot86}. Both models contain  nucleons interacting locally with 
meson fields. However, in contrast to the Walecka model, where the meson fields
are fundamental and the meson properties are introduced as phenomenological 
parameters, in the hybrid model the meson fields appears basically as 
$\bar qq$ pairs. 
Hence, we are able to calculate the meson masses and the physical coupling
constants for a given constituent mass in vacuum as well as in medium. In the 
latter case, there is a change of the chiral 
condensate which leads to a modification of the meson properties as 
well. Moreover, within the present model the nucleon can be described as a 
soliton, which allows for a check of the basic assumption of the model, namely
of the universality of the local coupling of the 
nucleons to the meson fields (\ref{nucmassv}).
  
As in the case of a quark medium, only the scalar field 
can have a non-zero expectation value in nucleon medium 
(constituent quark mass in medium) which is a 
solution of the corresponding equation of motion (\ref{medgap}) with a 
finite-temperature part written now in terms of nucleons
\begin{figure}[thb]
\input{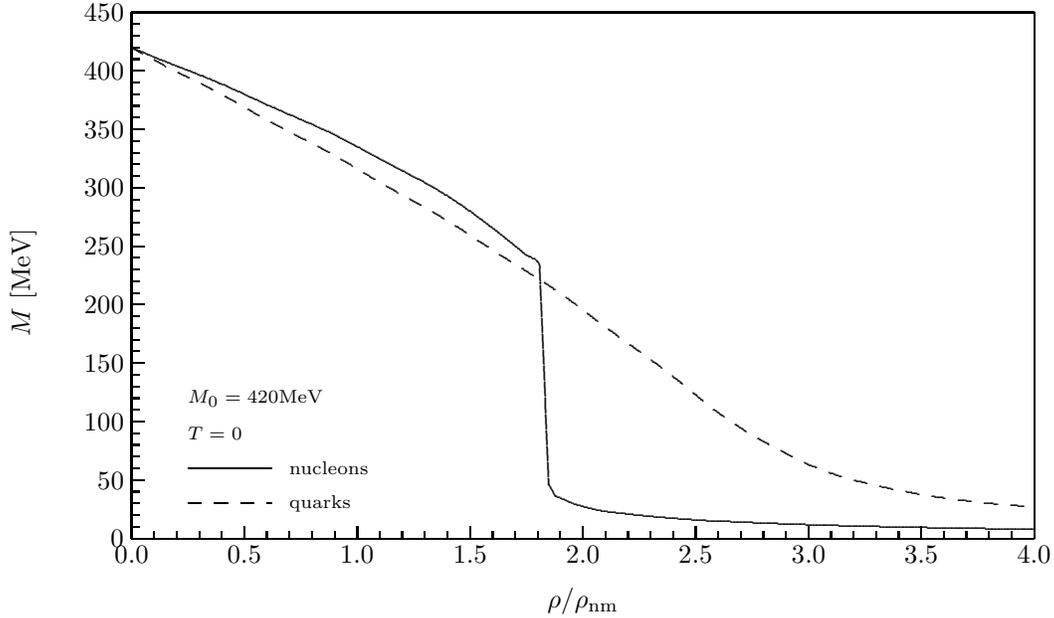}
\caption{Constituent quark mass $M$ as a function of density at $T=0$
in the quark (dashed line) and nucleon medium (solid line).}
\label{Figr7}
\end{figure}
\begin{equation}
N_c\sum_{\epsilon_n>0}{\bar\Phi_n\Phi_n
\over 1+\mbox{e}^{(\epsilon_n-\mu)\beta}}
- N_c\sum_{\epsilon_n<0}{\bar\Phi_n\Phi_n
\over 1+\mbox{e}^{-(\epsilon_n-\mu)\beta}}\longrightarrow
\sum_{\epsilon_n^N>0}{\bar\Phi_n^N\Phi_n^N
\over 1+\mbox{e}^{(\epsilon_n^N-\mu^N)\beta}}
-\sum_{\epsilon_n^N<0}{\bar\Phi_n^N\Phi_n^N
\over 1+\mbox{e}^{-(\epsilon_n^N-\mu^N)\beta}}
\label{medgapn} 
\end{equation}
together with the constraint $\rho_B=const$.

The results for $M$ are presented in fig.\ref{Figr6} at finite density as a 
function of $T$ and on fig.\ref{Figr7} as a function of density at $T=0$. At 
low $T$ values the curves are close to 
those of the quark matter which means that the use of a quark matter instead 
of nucleon one is a reasonable approximation which is not true for higher 
temperatures. Already at intermediate temperatures $T>120$ MeV they start to
deviate significantly. Whereas the quark matter is quite soft against thermal 
fluctuations, the nucleon matter is much stiffer and the corresponding 
chiral phase transition is rather sharp. Both the chiral condensate and the 
constituent quark mass show a discontinuity at the critical temperature. 
In fact, the system jumps between two
minima. The later is suggests that we have to deal with a first order 
phase transition even in the case of vanishing density in contrast to 
the case of quark matter where at temperatures $T>80$ MeV it is a 
second order one. The critical temperature is slightly larger than 
in the quark case and as we will see later it almost coincides with the 
critical 
temperature for the delocalization of the soliton which makes this model 
picture consistent. The results at $T=0$ and finite $\rho$ 
suggest a different model scenario. The critical density ($\approx$  
two times the nuclear matter density) for the chiral 
transition, which in fact, as we will see later, is also the critical density 
for the delocalization of the 
soliton in cold nucleon matter, is smaller than the one for the quarks 
matter. This model scenario suggests a delocalization 
transition from cold nucleon matter 
to quark one before the chiral phase transition to take place.

\begin{figure}[tb]
\input{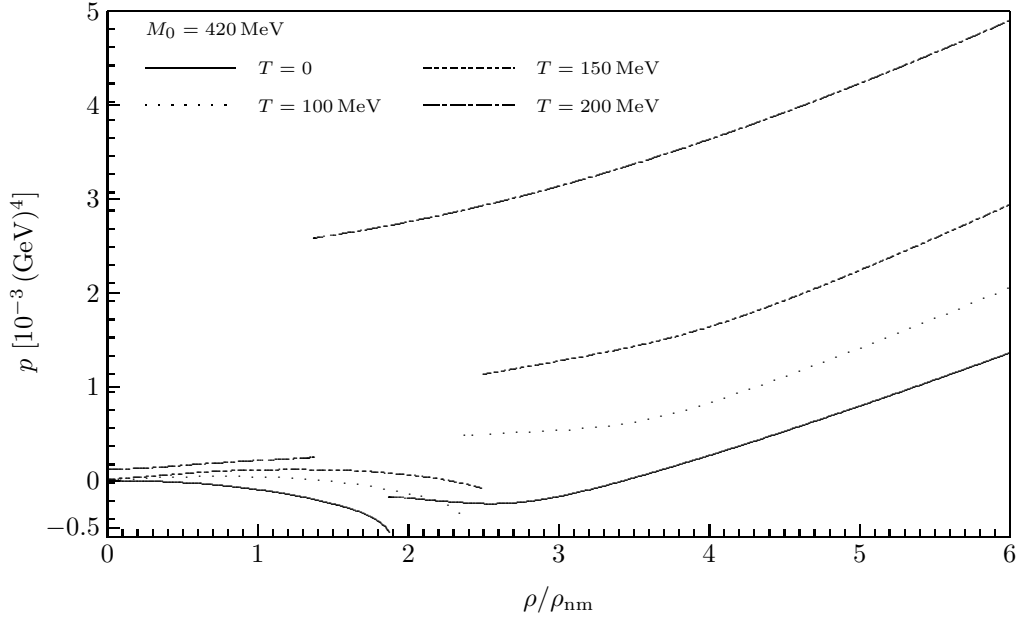}
\caption{EOS (pressure versus density) for different temperatures. The
pressure of the vacuum is subtracted. The discontinuities correspond
to a phase transition from a nucleon to a quark matter.}
\label{Figr8}
\end{figure}

\begin{figure}[hb]
\input{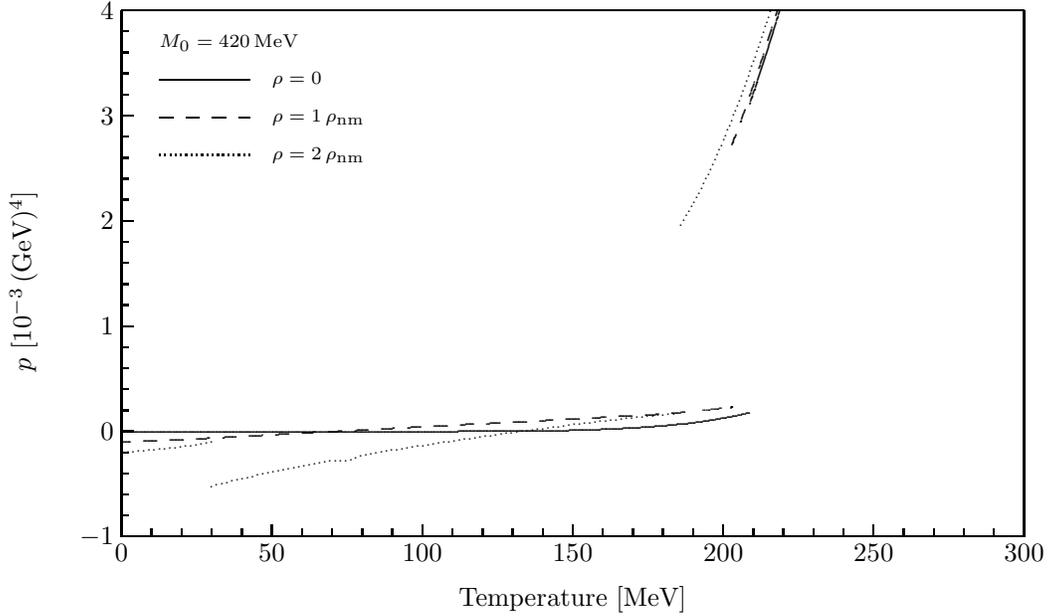}
\caption{Pressure as a function of temperature for different
densities. The discontinuities show the phase transition from the
nucleon to quark matter.}
\label{Figr9}
\end{figure}

\begin{figure}[thb]
\input{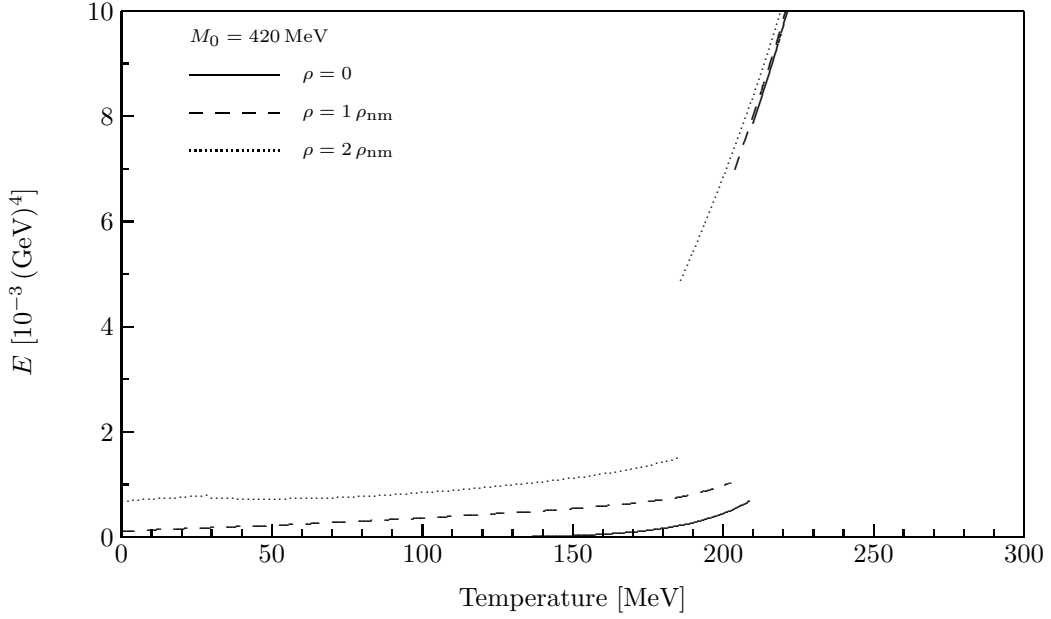}
\caption{Energy density as a function of temperature for different
densities. The discontinuities show the phase transition from the
nucleon to quark matter.}
\label{Figr10}
\end{figure}

According to the present model picture, at some critical temperature one 
expects a chiral phase transition together with a 
delocalization transition from nucleon matter to quark one. Apparently, it
would lead to drastic structural changes in the system. It can be clearly seen 
from the corresponding EOS (pressure versus baryon density) for different 
temperatures plotted in fig.\ref{Figr8}. On this figure 
we combine the results from the nucleon matter (below the critical
temperatures) in the hybrid model with those of the quark matter 
after the transition. 
Since we do not include vector mesons in the hybrid model, we are not able 
to reproduce the nuclear matter saturation at zero temperature and finite 
density which in the Walecka 
approach is due to the interplay between the $\sigma$-meson attraction and 
the $\omega$-meson repulsion. All curves in fig.\ref{Figr8} show rapid change
discontinuity at the 
delocalization transitions from nucleon to quark matter. Similar behavior can 
be found on the fig.~\ref{Figr9} and fig.~\ref{Figr10} for the pressure and 
energy density as a function of $T$ for different densities. 

Apart from the medium part written in terms of nucleons (\ref{medgapn}) 
the inverse meson propagators in the nucleon medium have the same structure 
(\ref{invprop}) as in the quark medium. In chiral limit the pions are massless
Goldstone bosons whereas the sigma mass is given by $2M$. Like the 
constituent mass $M$ the mass of the sigma meson in the nucleon 
medium has a discontinuity at the critical temperature. 

Our considerations are done in the mean-field approximation 
and suffer from the fact that meson quantum (loop) effects are not taken into 
account. However, as it is estimated 
in \cite{Hufner94,Zhuang94} the meson quantum effects may play a dominant role at low 
temperatures and vanishing density, whereas close to the critical temperature 
their contribution to the bulk thermodynamical variables is of order of 10~\% 
or even less. Hence, following the result of \cite{Hufner94,Zhuang94}, one 
expects
that the meson fluctuations would not change the model scenario for the 
phase transition presented above.    

\section{Nucleon as a soliton in medium}

As a next step we solve the $B=1$ soliton sector in a hot medium. We look for a
localized bound solution (soliton) of $N_c$ valence quarks interacting with 
the Fermi and Dirac sea both polarized due to the interaction.

\subsection{In quark medium}

First we consider the case of Fermi
and Dirac sea both filled with quarks of a constituent mass $M$. 
The thermodynamical potential (effective action) includes an explicit valence
quark contribution:
\begin{eqnarray}
\Omega(\mu,\beta) &=& N_c\theta(\epsilon_{val})\epsilon_{val}+ 
N_c\sum_{\epsilon_n<0} 
\Bigl[R^\Lambda_{3/2}(\epsilon_n)-R^\Lambda_{3/2}(\epsilon_n^0)
+(\mu-\epsilon_n)\Bigr]
- {N_c\over \beta} \sum_{\epsilon_n\neq val}
\ln\Bigl[1+\mbox{e}^{-\beta(\epsilon_n-\mu)}
\Bigr]\Bigr\} \nonumber\\
&+& {1\over V} \int_V \mbox{d}^3x \Biggl[\frac{1}{2G}(M^2-M_0^2)-\frac{m_0}{G} 
(\sigma-M_0)\Biggr] \,.
\label{regomega}
\end{eqnarray}
since the valence level is always occupied by $N_c$ quarks. 
 
The meson fields are assumed to be in a hedgehog form
\begin{equation}
\sigma({\bf r})=\sigma(r) \qquad \hbox{and} \qquad \vec\pi({\bf r})={\bf \hat
r}\pi(r) 
\end{equation}
and are restricted on the chiral circle
\begin{equation}
\sigma^2+{\vec\pi}^2=M^2\,. 
\label{cicircle}
\end{equation}
The latter is an {\it ad hoc} nonlinear constraint which
reflects the fact that the pions as almost massless Goldstone
bosons play a privileged role of dynamical mesons in this low-energy
region~\cite{Diakonov88}.

From (\ref{statcond}) one gets the equations of motion for the meson fields
\begin{eqnarray}
\sigma(\vec r)&=&-\,G\,N_c\Biggl\{\sum_{n}
\bar\Phi_n(\vec r)\Phi_n
(\vec r)R_{1/2}^\Lambda(\epsilon_n)\epsilon_n+
\theta(\epsilon_{val})\bar\Phi_{val}(\vec r)\Phi_{val}
(\vec r)+\sum_{\epsilon_n>0}{\bar\Phi_n(\vec r)\Phi_n(\vec r)
\over 1+\mbox{e}^{(\epsilon_n-\mu)\beta}}\nonumber\\
&-&\sum_{\epsilon_n<0}{\bar\Phi_n(\vec r)\Phi_n(\vec r)
\over 1+\mbox{e}^{-(\epsilon_n-\mu)\beta}}\Biggr\}+m_0\, 
\label{sigmeseq}
\end{eqnarray}
\begin{figure}[thb]
\input{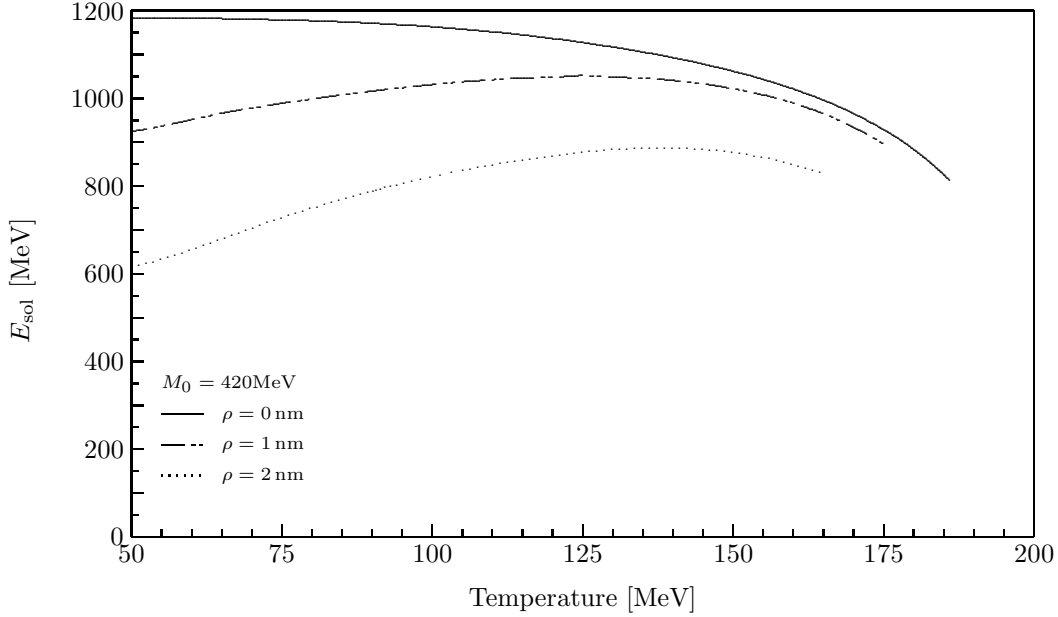}
\caption{Soliton energy as a function of the temperature at
three different quark matter densities.}
\label{Figr11}
\end{figure}
and 
\begin{eqnarray}
\pi(\vec r)&=&-\,G\,N_c\Biggl\{\sum_{n}
\bar\Phi_n(\vec r) i\gamma_5({\bf \hat r}\cdot \vec \tau)\Phi_n(\vec r)
R_{1/2}^\Lambda(\epsilon_n)+
\theta(\epsilon_{val})\bar\Phi_{val}(\vec r)i\gamma_5({\bf \hat r}
\cdot \vec \tau) \Phi_{val} (\vec r)\nonumber\\
&+&\sum_{\epsilon_n>0}{\bar\Phi_n(\vec r)i\gamma_5
({\bf \hat r}\cdot \vec \tau)\Phi_n(\vec r)
\over 1+\mbox{e}^{(\epsilon_n-\mu)\beta}}
-\sum_{\epsilon_n<0}{\bar\Phi_n(\vec r)i\gamma_5
({\bf \hat r}\cdot \vec \tau)\Phi_n(\vec r)
\over 1+\mbox{e}^{-(\epsilon_n-\mu)\beta}}\Biggr\}\,. 
\label{pimeseq}
\end{eqnarray}
We use a numerical self-consistent iterative procedure based on a method
proposed by Ripka and Kahana~\cite{Kahana84}. The procedure consists in
solving in an iterative way the Dirac equation (\ref{dirac}) together with the
equations of motion of the meson fields - eqs.(\ref{sigmeseq}),(\ref{pimeseq}),
and the constraint $N_B=const$ (\ref{barnumber}) which is actually a 
condition fixing the chemical potential $\mu$.  

The techniques for the numerical procedure
are well known~\cite{Reinhardt88,Meissner89} for the case
of vanishing chemical potential where there is no contribution coming from
the positive continuum. They can be easily adopted to the case of finite
chemical potential and temperature where an 
additional source (Fermi sea quarks) for the meson fields appears in 
eqs.(\ref{sigmeseq}),(\ref{pimeseq}). 

In the case of fixed $T$ and $\rho$ the proper way to describe the equilibrium
state of a thermodynamical system is to use the free energy (\ref{freeenergy}).
Hence, the energy of the $B=1$ soliton (effective soliton mass) is given by 
the change of the free energy when the $N_c$ valence quarks are added to the 
medium. Apparently, in this the baryon number of the system increases by 
one, $N_B+1$.
Since due to the presence of the soliton both the Fermi sea and the Dirac sea 
are getting polarized, subtracting the free energy $F(\mu_0,\beta)$ of the
unperturbed Fermi and Dirac sea (translationally invariant medium
solution), the soliton energy is given
by the sum of the energy of the valence quarks and
the contributions due to the polarization of both continua:
\begin{eqnarray}
E_{sol}\,&=&\, N_c\eta_{val}\epsilon_{val} + N_c\sum_{\epsilon_n<0} 
\Bigl[R^\Lambda_{3/2}(\epsilon_n)+(\mu-\epsilon_n)\Bigr]
- {N_c\over \beta} \sum_{\epsilon_n\neq val} 
\ln\Bigl[1+\mbox{e}^{-\beta(\epsilon_n-\mu)}\Bigr] \nonumber\\
&-& {1\over V} \int_V \mbox{d}^3x {m_0\over G} \sigma+\mu N_c\rho_B
-F(\mu_0,\beta)\,.
\label{solenergy}
\end{eqnarray}
\begin{figure}[thb]
\input{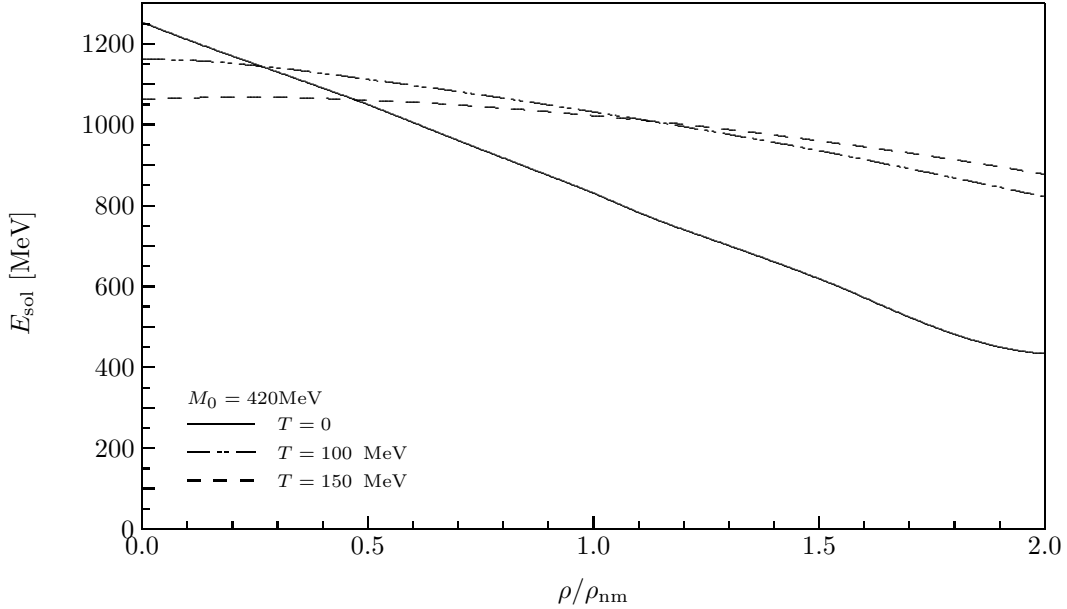}
\caption{Soliton energy as a function of density for vanishing
as well as finite temperature values.}
\label{Figr12}
\end{figure}
The chemical potential $\mu_0$ fixed by the constrain $N_B=const$ 
(\ref{barnumber}) corresponds to the translationally invariant medium 
solution. The soliton baryon density distribution is split in valence, sea 
and medium parts:
\begin{eqnarray}
\rho_{sol}(\vec r)&=&\theta(\epsilon_{val})
\Phi_{val}^\dagger(\vec r)\Phi_{val}(\vec r)+{1\over2}\sum_{\epsilon_n}
\Phi_n^\dagger(\vec r)\Phi_n(\vec r)\mbox{sgn}(-\epsilon_n)
\nonumber\\
&+&\sum_{\epsilon_n>M}{\Phi_n^\dagger(\vec r)\Phi_n(\vec r)
\over 1+\mbox{e}^{(\epsilon_n-\mu)\beta}}
-\sum_{\epsilon_n<0}{\Phi_n^\dagger\Phi_n\over 
1+\mbox{e}^{-(\epsilon_n-\mu)\beta}}-\rho_B \,.
\label{soldensity}
\end{eqnarray}
Because of the constraint $N_B=const$, only the first term in r.h.s. of
(\ref{soldensity}) contributes to the baryon number $B=1$. The soliton 
m.s.radius defined as
\begin{equation}
<r^2>_{sol}=\int \mbox{d}^3r r^2\rho_{B=1}(\vec r)\,
\label{msradius}
\end{equation}
can be used to measure the spatial extension of the soliton.

Our results for the soliton energy as a function of the temperature for 
vanishing as well as for finite density values are presented in 
fig.~\ref{Figr11}. For completeness on the fig.~\ref{Figr12} we also present 
the soliton energy as a function of density at vanishing~\cite{Christov93} 
as well as at finite temperature. 

We do not find a localized solution (soliton) at
temperatures larger than a critical value $T_{cr}^B\approx 180$ MeV. 
It means that at large enough values 
the temperature effects simply disorder the system and destroys the 
soliton. Within the present model picture this may be interpreted as an 
indication for a delocalization of the nucleon in hot medium but, 
however, one should keep in mind that the model lacks confinement.
It should be mentioned that a similar effect has been found in the 
Skyrme model~\cite{Bernard89}.

As can be seen from fig.~\ref{Figr11} the temperature effects for the soliton
are much weaker than the finite
density effects. In the case of finite $T$ and zero 
$\rho$ the soliton energy shows almost no reduction for temperatures not 
close to the critical one, whereas in the opposite case of finite $\rho$ and  
zero $T$ the soliton energy is linearly decreasing with the density. 
In the hot medium (both 
$T$ and $\rho$ finite) the temperature clearly suppresses the finite density 
effects 
and stabilizes the soliton. The latter results in a non-monotonic temperature 
dependence of the soliton energy at fixed finite density. At densities 
larger than the nuclear matter one $\rho_{nm}$, which are relevant for the 
heavy-ion experiments, the reduction of the soliton energy at finite 
temperature is much smaller than one at $T=0$ (see
fig.~\ref{Figr12}). 
\begin{figure}[thb]
\input{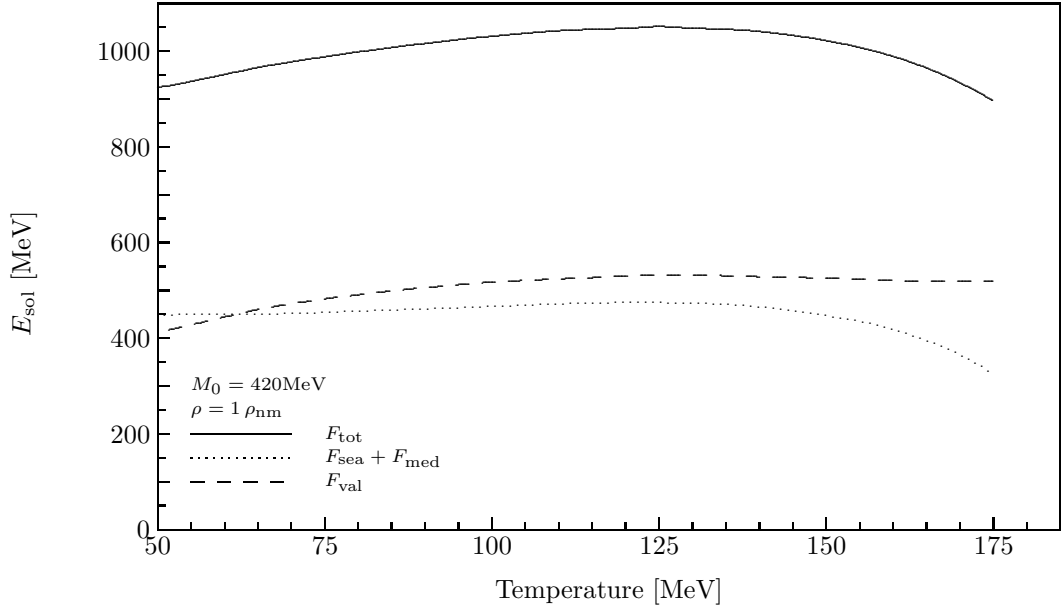}
\caption{Soliton energy and contributions coming from the
valence quarks and from the polarized continua as a function of
temperature at density $\rho = \rho_{\rm nm}$.}
\label{Figr13}
\end{figure}

In 
particular, at $\rho_{nm}$ and $T> 100$ MeV the reduction is about 15\% 
whereas at $T=0$ is two times larger. Further, at densities larger than two 
times 
$\rho_{nm}$ the soliton exists only at intermediate temperatures 
$100$~MeV~$<T<180$~MeV. It means that according to the present model 
calculations, the soliton is 
more stable in the hot matter than in the cold one. In fact, it can be 
easily understood. Because of the gap in the quark spectrum, $2M$, the Dirac 
sea is much less affected by the 
temperature than the Fermi sea (positive-energy part of the spectrum). 
Disordering mainly the Fermi sea the attractive interaction between the medium 
and the valence quarks and Dirac sea, which destabilize the soliton, is 
diminished. 
\begin{figure}[bht]
\input{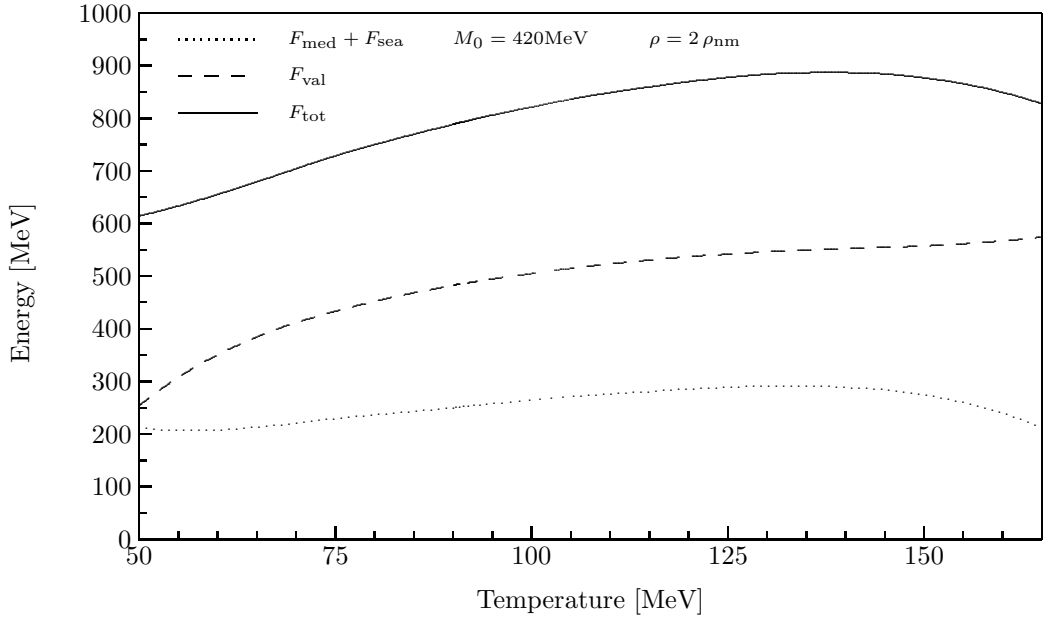}
\caption{Soliton energy and contributions coming from the
valence quarks and from the polarized continua as a function of
temperature at density $\rho = 2\,\rho_{\rm nm}$.}
\label{Figr14}
\end{figure}
At temperatures close to the critical one, the thermal 
fluctuations become comparable with the chiral order parameter, the chiral 
condensate $\bar qq>$, they completely disorder the system and in particular,
destroy the soliton.  

It is interesting to look at the separated valence and sea contributions to 
the soliton energy presented in fig.~\ref{Figr13} and \ref{Figr14} for two 
different medium densities, namely one and two times nuclear matter density, 
respectively. Whereas at low temperature
almost a half of the energy comes from the polarized Dirac and Fermi sea, at 
increasing $T$ the valence contribution becomes dominant. Also, in contrast to 
the valence quark contribution, which shows a strong dependence on $T$, 
the polarized sea contribution stays almost constant with $T$ not close to 
the critical one but it is strongly affected by the medium density. In 
particular, going from one to two times $\rho_{nm}$, the sea contribution
is reduced by a factor of two, which destabilizes the soliton. Further, at 
$T$ close to the critical values it vanishes.   

\begin{figure}[bht]
\input{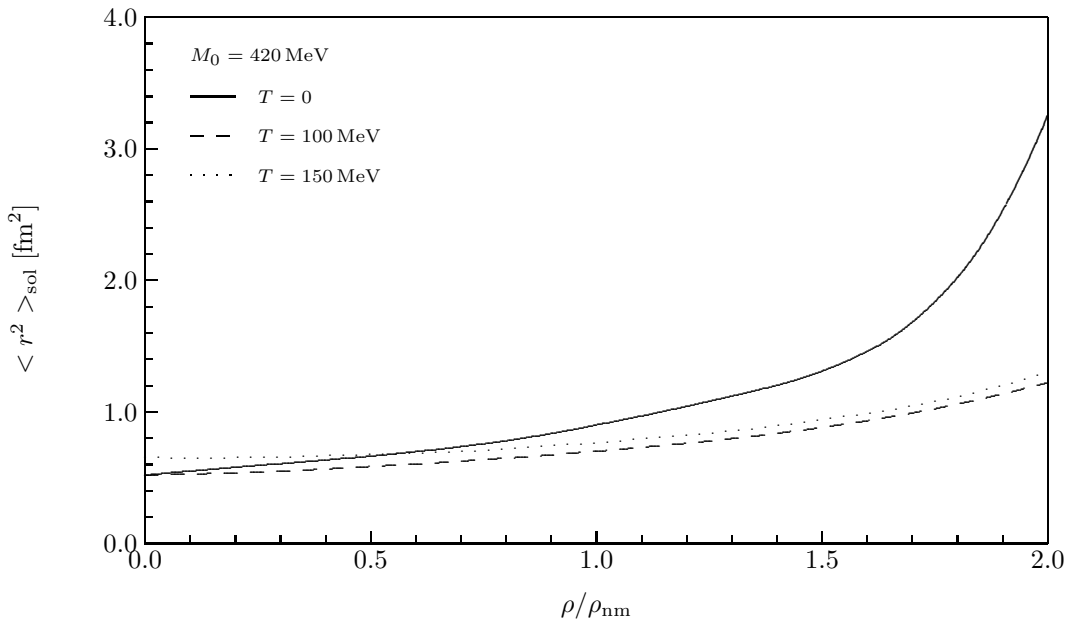}
\caption{Soliton m.s. radius as a function of density for
vanishing as well as for finite temperature values.}
\label{Figr15}
\end{figure}

In order to illustrate the change of the soliton structure in hot medium  we 
also plot the soliton square radius (\ref{msradius}) as a function of 
density for different temperatures in figs.~\ref{Figr15}.  
All curves in fig.~\ref{Figr15} show a clear trend to grow rapidly at 
temperatures close to the 
critical values which is an indication for delocalization of the soliton.
At finite both density and temperature the radius is smaller than in the case of cold matter which is a sign for a stabilization of the soliton in hot 
medium compared to the case of cold one. At densities larger than 
two times $\rho_{nm}$ the soliton exists only at intermediate temperatures.  

\subsection{In medium of nucleons}

In this subsection we consider the $B=1$ soliton in the hybrid model 
with $N_c$ valence quarks interacting with both the quark Dirac sea and the 
nucleon Fermi sea via the meson fields.
As in the quark medium the $B=1$ soliton is obtained solving the Dirac 
equations~(\ref{dirac}),(\ref{ndirac}) together with meson equations of motion 
(\ref{sigmeseq}),(\ref{pimeseq}) which contain the finite-temperature nucleon 
part (\ref{medgapn}), and the constraint $\rho_B=const$ (\ref{bardensnucl}) to 
fix the chemical potential $\mu^N$ in the self-consistent iterative procedure. 
The energy of the $B=1$ soliton is given again by the change of
the free energy when the $N_c$ valence quarks are added to the system. 

\begin{figure}[thb]
\input{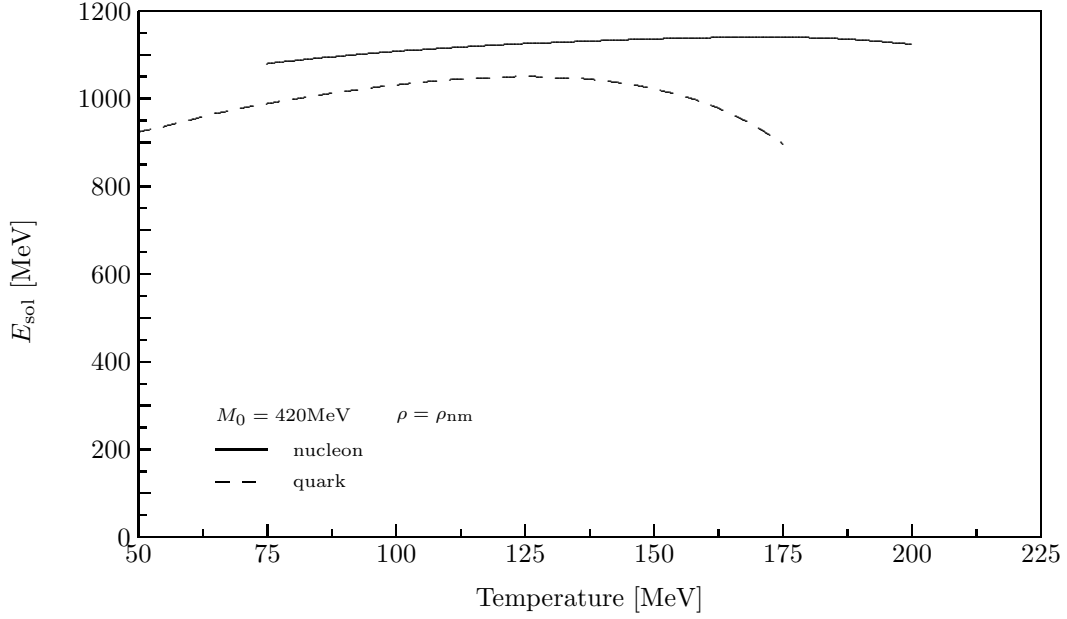}
\caption{Soliton energy in a nucleon matter compared to those in a
quark matter for $\rho = \rho_{\rm nm}$.} 
\label{Figr16}
\end{figure}

\begin{figure}[bht]
\input{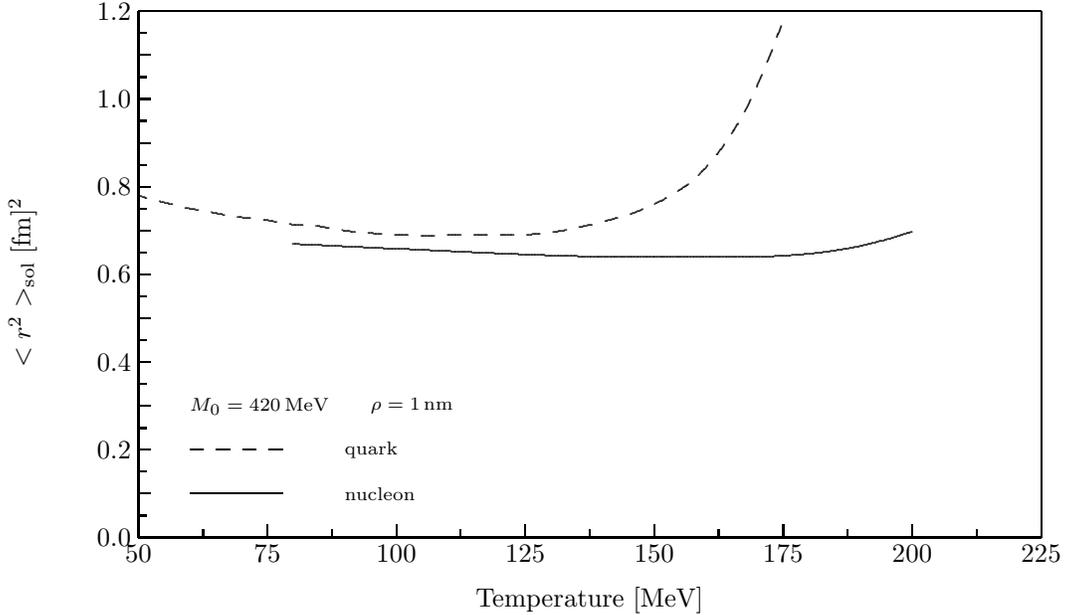}
\caption{Soliton m.s. radius in a nucleon matter compared to those in
a quark matter for $\rho = \rho_{\rm nm}$.} 
\label{Figr17}
\end{figure}

On fig.\ref{Figr16} the temperature dependence of the calculated $B=1$ soliton 
energy for the nuclear matter density is compared with those of the quark 
matter.  The two curves have similar trends at intermediate values of $T$ and 
start to deviate at $T$ close to critical one. The soliton in the nucleon 
matter is more bound and less affected by the temperature. At some 
critical temperature of about 200 MeV the soliton disappears -- delocalization
of the soliton. In fact, at the same temperature in the hybrid model there is 
also the chiral phase transition from Goldstone to 
Wigner phase. The m.s. soliton radius is presented in fig.\ref{Figr17}. 
Close to the critical temperature it starts to grow  
but it is much less pronounced than in the case of quark matter.

\begin{figure}[htb]
\input{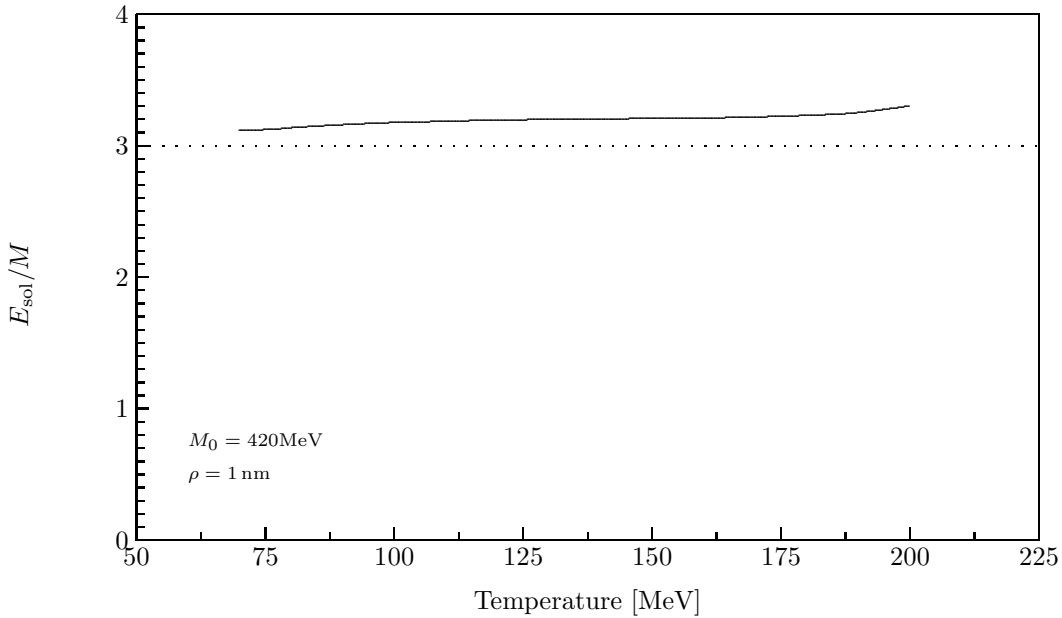}
\caption{Ratio $E_{\rm sol}/M$ in a nucleonic medium for $\rho =
\rho_{\rm nm}$ as a function of temperature.}
\label{Figr18}
\end{figure}

Back to our discussion concerning the Walecka model, in our case this model 
picture 
is valid to some critical density of about two times nuclear matter density. 
At higher densities we found a delocalization transition to a quark matter. On 
fig.\ref{Figr18} we also plot the coupling constant 
\begin{equation}
g_N=E_{sol}/M
\label{gN}
\end{equation}
as a function of $T$ at one nuclear matter density. As can be seen it stays 
almost constant which means that the relation (\ref{nucmassv}) is a good 
approximation also in hot medium.

 In contrast to the quark matter the energy 
of the soliton in the 
nucleon matter is slightly above the threshold of $3M$. However, in the hybrid 
model this channel is physically closed by construction - before the phase 
transition, in the Goldstone phase, free quarks are not allowed to occupy the 
positive part of the quark spectrum.  

\section{Summary}

We study the bulk thermodynamical properties, some meson properties and the 
nucleon as a $B=1$ soliton in hot medium using the NJL model in mean-field
approximation. We consider the case of a quark medium in the NJL 
model as well as of a  
nucleon medium  in a hybrid NJL model in which the quark Dirac sea is 
combined with a nucleon Fermi sea. In both cases we find that 
the chiral order parameter, the chiral condensate, gets modified and at some 
critical values of temperature and/or density vanishes which indicates a 
chiral phase transition from Goldstone to Wigner phase. At finite density the 
chiral order 
parameter and the constituent quark mass have a non-monotonic temperature 
dependence and at finite temperatures not close to the critical one they are 
less affected than in the cold matter. Below the chiral phase transition the 
pions are Goldstone bosons and their mass remains almost unchanged whereas the 
sigma mass follows the constituent quark mass. After the phase transition the 
mesons are parity-doubled and 
 appear as broad resonances in the $\bar qq$ continuum. With increasing 
temperature the position of the resonance moves to higher energies and and its 
width increases very fast almost independently of the density. At some 
temperature of about 300 MeV the meson structure is very weak and is almost 
washed out. Whereas at low density
and temperature values the quark matter and the nucleon matter provide similar
results, at larger temperatures they differ significantly. In particular, the 
quark matter is much softer against thermal fluctuations and the chiral phase 
transition is rather smooth. The nucleon matter is much less affected by the 
temperatures even for values close to the critical one. The chiral phase 
transition is rather sharp and all thermodynamical variables show a large 
discontinuities which is an indication for a first order phase transition. 
We study also the structure of the 
baryon number one soliton of the NJL model immersed in a hot quark as well as 
in a nucleon medium. The polarization of both the Fermi and the Dirac sea is 
taken into account in a consistent way. We find that at finite density the 
temperature stabilizes the soliton and after some critical density it exists 
only at intermediate temperature values. In general, at finite temperature 
the soliton is 
more bound and less swelled than in the case of a cold matter. At some critical
temperature we find no more a localized solution. In the case of nucleon 
matter this delocalization means a transition from the nucleon to the quark 
matter. The critical temperature
coincides with those for the chiral phase transition. According to this model 
scenario one should expects at some critical temperature a common sharp phase 
transition from the nucleon to the quark matter with a restoration of chiral 
symmetry. All present results are obtained in mean-field (large $N_c$) 
approximation  which means that the meson quantum (loop) effects are not 
included. However, it is 
expected~\cite{Hufner94,Zhuang94} that these effects would play a minor role 
near the phase transition and hence it would not change principally our 
results.

{\large \bf Acknowledgements}:
The work has been supported partially by the Bundesministerium f\"ur Bildung
und Wissenschaft and the Bulgarian Science Foundation under contract 
$\Phi$--32.

\end{document}